\documentclass{article}
\usepackage[latin9]{inputenc}
\usepackage{color}
\usepackage{amsmath}
\usepackage{amssymb}
\usepackage{graphicx}
\usepackage{esint}
\usepackage{hyperref}
\usepackage{soul}

\makeatletter
\usepackage{sidecap}\usepackage{float}\usepackage{setspace}\usepackage{epstopdf}\usepackage{comment}\usepackage{color}\usepackage[font={small,it}]{caption}

\@ifundefined{showcaptionsetup}{}{%
 \PassOptionsToPackage{caption=false}{subfig}}
\usepackage{subfig}
\makeatother

\begin{document}
\date{}
\title{Lagrangian Coherent Structures from Video Streams of Jupiter\footnote{\textcolor{red}{Publication in the MFO report}}}

\author{Alireza Hadjighasem \and\
George Haller}
\maketitle

\section{Introduction}
 Jupiter's fast rotation - one rotation over 10 hours - creates strong jet streams, smearing
its clouds into linear bands of dark and light zonal belts that circle
the planet on lines of almost constant latitude. Such a high degree
of axisymmetry is absent in our own atmosphere. Moreover, Jupiter
has the largest and longest-living known atmospheric vortex, the Great
Red Spot (GRS). Such vortices abound in nature, but GRS's size, long-term persistence, and temporal
longitudinal oscillations make it unique.

Here, we uncover, for the first time, unsteady material structures
that form the cores of zonal jets and the boundary of the GRS in Jupiter's
atmosphere. We perform our analysis on a velocity field extracted
from a video footage acquired by the NASA Cassini spacecraft.

\section{Background: Lagrangian coherent structures\label{sec:Set-up}}

Consider a two-dimensional unsteady velocity field 
\begin{equation}
\dot{x}=v(x,t),\quad x\in U\subset\mathbb{R}^{2},\quad t\in[t_{0},t],\label{eq:dynsys}
\end{equation}
which defines a two-dimensional flow over the finite time interval
$[t_{0},t]$ in the spatial domain $U$. The flow map $F_{t_{0}}^{t}(x_{0}):x_{0}\mapsto x_{t}$
of \eqref{eq:dynsys} then maps the initial condition $x_{0}$ at
time $t_{0}$ to its evolved position $x_{t}$ at time $t$. The Cauchy--Green
(CG) strain tensor associated with \eqref{eq:dynsys} is defined as
\begin{equation}
C_{t_{0}}^{t}(x_{0})={DF_{t_{0}}^{t}}^{\intercal}DF_{t_{0}}^{t},
\end{equation}
where $DF_{t_{0}}^{t}$ denotes the gradient of the flow map, and
the symbol $\intercal$ indicates matrix transposition. The CG strain
tensor is symmetric and positive definite, thus has two positive eigenvalues
$0<\lambda_{1}\leq\lambda_{2}$ and an orthonormal eigenbasis $\left\{ \xi_{1},\xi_{2}\right\} $,
defined as 
\begin{align}
 & C_{t_{0}}^{t}(x_{0})\xi_{i}(x_{0})=\lambda_{i}(x_{0})\mathbf{\xi}_{i}(x_{0}),\quad\left|\xi_{i}(x_{0})\right|=1,\quad i\in\{1,2\},
\end{align}
We shall suppress the dependence of CG invariants on $t_{0}$
and $t$ for notational simplicity.\\
A general material line (composed of an evolving curve of initial conditions) experiences both shear and strain in its deformation.
As argued in \cite{faraz,black_hole}, the averaged straining and
shearing experienced within a strip of $\epsilon$-close material
lines will generally differ by an $\mathcal{O\left(\epsilon\right)}$
amount over a finite time interval due to the continuity of the finite-time
flow map.\\
 We seek a Lagrangian Coherent Structure (LCS) as an exceptional material
line around which $\mathcal{O\left(\epsilon\right)}$ material belts
show no $\mathcal{O\left(\epsilon\right)}$ variation in the length-averaged
Lagrangian shear or strain over the time interval $[t_{0},t]$. This
implies that an LCS is a stationary curve for the averaged Lagrangian
shear or strain functionals.

\subsection{Shearless LCSs: stationary curves of averaged shear}

Specifically, a shearless LCS is a material line whose averaged shear
shows no leading order variation with respect to the normal distance
from the LCS. Farazmand et al. \cite{faraz} show that such curves are null-geodesics
of a Lorentzian metric. The most robust class of these null-geodesics turns out to be composed of
smooth chains of tensorlines (trajectories of the eigenvector fields
of CG) that connect singularities of the CG field. Out of all such
possible chains, one builds parabolic LCSs (generalized jet cores)
by identifying tensorlines closest to being neutrally stable (cf.
Farazmand et al. (2014) for details). 

\subsection{Strainless LCSs: stationary curves of averaged strain}

Similarly, a strainless LCS is a material line whose averaged strain
shows no leading order variation with respect to the normal distance
from the LCS. As shown by Haller and Beron--Vera \cite{black_hole}, such stationary
curves of the tangential stretching functional coincide with the null-geodesics
of another Lorentzian metric that are tangent to one of the vector fields 
\begin{equation}
\eta_{\lambda}^{\pm}(x_{0})=\sqrt{\frac{\lambda_{2}(x_{0})-\lambda^{2}}{\lambda_{2}(x_{0})-\lambda_{1}(x_{0})}}\xi_{1}(x_{0})\pm\sqrt{\frac{\lambda^{2}-\lambda_{1}(x_{0})}{\lambda_{2}(x_{0})-\lambda_{1}(x_{0})}}\xi_{2}(x_{0}),\label{eta_vector}
\end{equation}
We refer to the closed
orbits (limit cycles) of the vector fields \eqref{eta_vector} as
elliptic LCS. The outermost orbit of such a family of limit cycles
serves as a coherent Lagrangian vortex boundary. It is infinitesimally
uniformly stretching, i.e., any of its subsets stretches exactly by
a factor of $\lambda$ over the time interval $[t_{0},t]$. Limit
cycles of $\eta_{\lambda}^{\pm}(x_{0})$ only tend to exist for $\lambda\approx1$,
guaranteeing a high degree of material coherence for the Lagrangian
vortex boundary.

\section{Results}

We used the ACCIV algorithm of Asay-Davis et al. \cite{ACCIV} to extract
a time-resolved atmospheric velocity field from video footage taken
by the NASA Cassini Orbiter. The ACCIV algorithm yields a high density
of wind velocity vectors, which is advantageous over the limited number
of vectors traditionally obtained from manual cloud tracking.\\
 Observational records of Jupiter go back to the late 19th century,
indicating that Jupiter's atmosphere is highly stable in the latitudinal
direction. Therefore, the average zonal velocity profile as a function
of latitudinal degree is an important benchmark for examining the
quality of the reconstructed velocity field. 

\begin{figure}
    \subfloat[\label{fig:vel_comparison}]
    {{\includegraphics[width=0.48\textwidth]{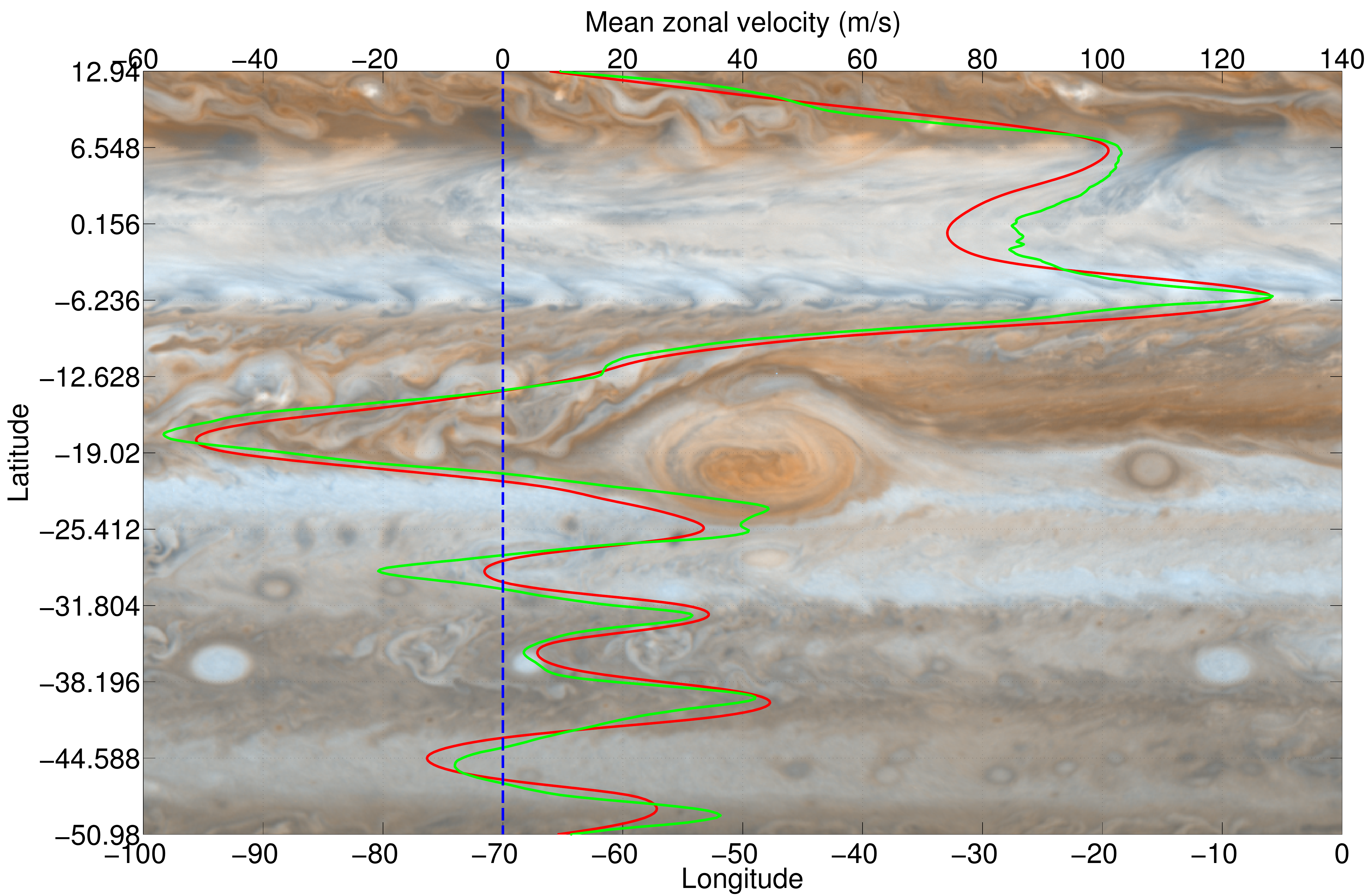} }}
    \subfloat[\label{fig:full_map}]
    {{\includegraphics[width=0.54\textwidth]{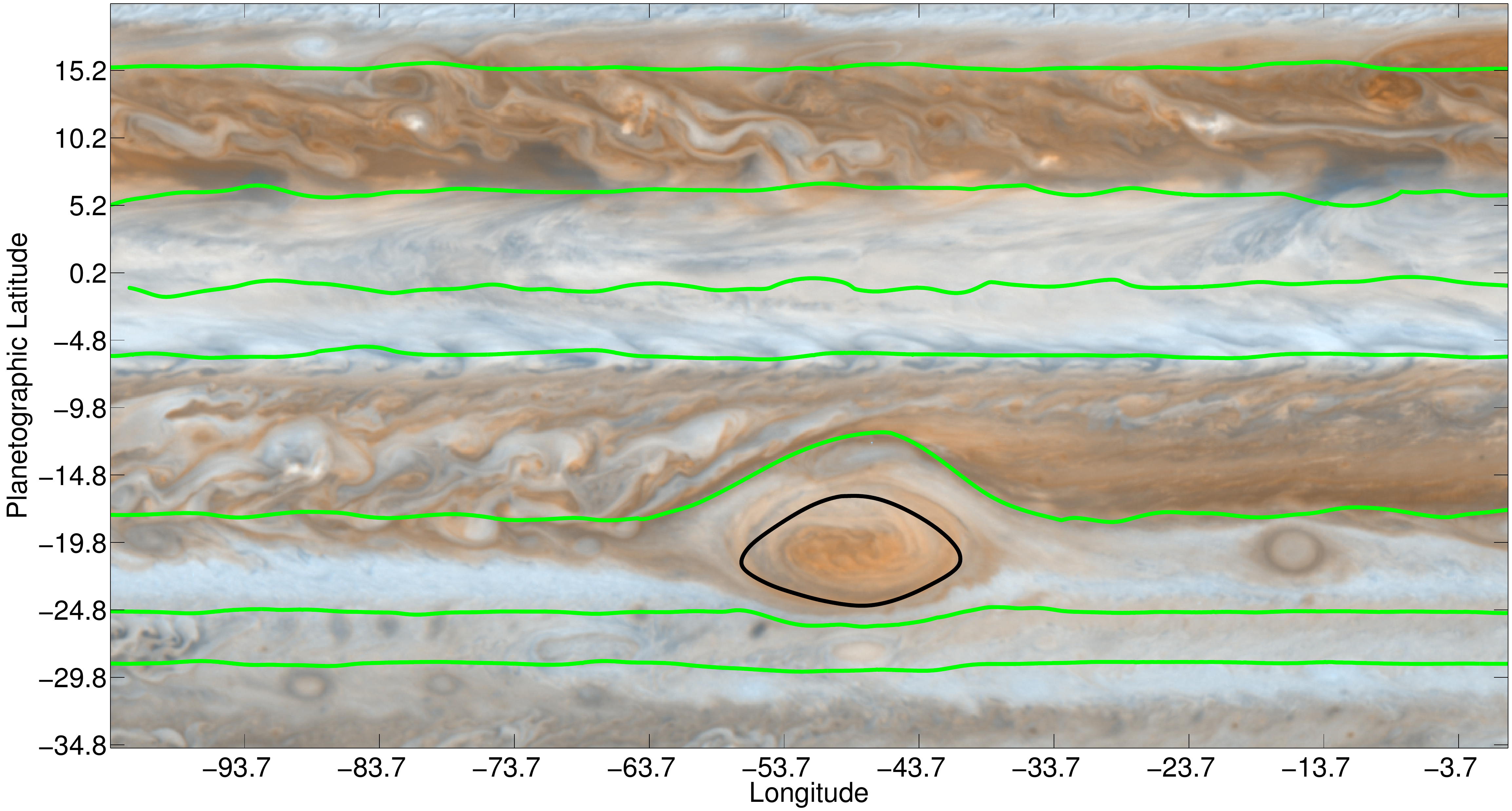} }}
    \caption{(A) Zonal velocity profile of Jupiter's atmosphere. The red line is our velocity profile which is obtained from Cassini images, and the green profile is the velocity profile reported by S.S. Limaye \cite{Limaye}. (B) Shearless LCS (green) as zonal jet cores, and an elliptic LCS (black) as the Lagrangian boundary of the Great Red Spot. (Background image credit: NASA/JPL-Caltech)}%
    \label{fig:example}%
\end{figure}

In Fig. \ref{fig:vel_comparison}, we compare our temporally averaged
zonal velocity profile obtained form ACCIV algorithm with the profile
reported by Limaye \cite{Limaye}. Limaye's profile is based on Voyager I
and Voyager II images, covering 144 Jovian days. Our velocity profile
is based on video footage captured by Cassini Orbiter during its flyby
in route to Saturn in 2000, just covering 24 Jovian days. Despite
these differences in the data, the two profiles show sufficiently
close agreement.\\
Using the extracted time-resolved velocity, we applied the geodesic
theory of LCSs reviewed in Section \ref{sec:Set-up} to the detection
of a coherent Lagrangian boundary for the GRS, and of the cores of
eastward- and westward-moving zonal jets (Fig. \ref{fig:full_map}).
Advected images (not shown here) of the extracted LCSs confirm their
sustained coherence and organizing role in cloud transport and mixing.
We will report further details and results elsewhere.

%\pagebreak
\renewcommand{\section}{\subsubsection}
\bibliographystyle{plain} 

\end{document}